\documentclass{article} 

\usepackage{amsmath,amsfonts,bm}

\def\1{\bm{1}}

\DeclareMathAlphabet{\mathsfit}{\encodingdefault}{\sfdefault}{m}{sl}
\SetMathAlphabet{\mathsfit}{bold}{\encodingdefault}{\sfdefault}{bx}{n}

\newcommand{\E}{\mathbb{E}}

\newcommand{\R}{\mathbb{R}}

\DeclareMathOperator*{\argmin}{arg\,min}

\usepackage[hidelinks]{hyperref}
\usepackage{url}
\usepackage{graphicx}
\usepackage{amssymb}
\usepackage{algorithm}
\usepackage{algpseudocode}
\usepackage{multirow}
\usepackage{adjustbox}
\usepackage{tabularx,booktabs}
\usepackage{mathtools}
\usepackage{amsthm}
\usepackage{wrapfig}
\usepackage{caption}
\usepackage{xcolor,colortbl}
\usepackage{natbib}
\usepackage{fullpage}

\newcommand{\best}[1]{%
	{\textbf{\color[HTML]{D52815}#1}}%
}
\newcommand{\secondbest}[1]{%
	{\underline{\color[HTML]{00008A}#1}}%
}

\definecolor{lightgreen}{rgb}{.9,1,.9}
\definecolor{darkblue}{rgb}{0,0 ,0.542}

\hypersetup{
    colorlinks = true,
    citecolor = darkblue,
    linkcolor = {black}
}

\theoremstyle{plain}
\newtheorem*{theorem*}{Theorem}

\newtheorem{theorem}{Theorem}

\newtheorem{assumption}{Assumption}

\def\defn{\,\coloneqq\,}

\def\d{{\mathsf{\, d}}}

\def\log{{\mathsf{log}}}

\def\min{\mathop{\mathsf{min}}}

\def\R{\mathbb{R}}
\def\E{\mathbb{E}}

\def\ebm{{\bm{e}}}

\def\xbm{{\bm{x}}}
\def\zbm{{\bm{z}}}
\def\ybm{{\bm{y}}}
\def\zbm{{\bm{z}}}
\def\sbm{{\bm{s}}}

\def\bbm{{\bm{b}}}

\def\nbm{{\bm{n}}}

\def\kbm{{\bm{k}}}

\def\nablahat{{\widehat{\nabla}}}

\def\Abm{{\bm{A}}}
\def\Dbm{{\bm{D}}}
\def\Pbm{{\bm{P}}}
\def\Fbm{{\bm{F}}}
\def\Sbm{{\bm{S}}}

\def\Kbm{{\bm{K}}}
\def\Wbm{{\bm{W}}}

\def\thetabm{{\bm{\theta }}}

\def\Hbf{{\mathbf{H}}}
\def\Ibf{{\mathbf{I}}}

\def\Abm{{\bm{A}}}
\def\Dbm{{\bm{D}}}
\def\Pbm{{\bm{P}}}
\def\Fbm{{\bm{F}}}

\def\Mbm{{\bm{M}}}

\def\Ncal{{\mathcal{N}}}

\def\Isf{{\mathsf{I}}}
\def\Rsf{{\mathsf{R}}}
\def\Tsf{{\mathsf{T}}}

\def\Tsf{{\mathsf{T}}}

\def\Isf{{\mathsf{I}}}

\def\xbmhat{{\widehat{\bm{x}}}}

\def\argmin{\mathop{\mathsf{arg\,min}}} 

\newcommand{\norm}[1]{\left\lVert#1\right\rVert}

\title{Stochastic Deep Restoration Priors for Imaging Inverse Problems}

\author{Yuyang Hu\(^1\),  Albert Peng\(^1\), Weijie Gan\(^1\), \\ {Peyman Milanfar}\(^2\), {Mauricio Delbracio}\(^2\), Ulugbek S. Kamilov\(^{1}\) \\
\(^1\)Washington University in St. Louis, \(^2\)Google\\
\texttt{\{h.yuyang, albertpeng, weijie.gan, kamilov\}@wustl.edu},\\ \texttt{\{milanfar,mdelbra\}@google.com}  \\
}

\begin{document}
\date{}

\maketitle
\begin{abstract}
Deep neural networks trained as image denoisers are widely used as priors for solving imaging inverse problems. While Gaussian denoising is thought sufficient for learning image priors, we show that priors from deep models pre-trained as more general restoration operators can perform better. We introduce \emph{Stochastic deep Restoration Priors (ShaRP)}, a novel method that leverages an ensemble of such restoration models to regularize inverse problems. ShaRP improves upon methods using Gaussian denoiser priors by better handling structured artifacts and enabling self-supervised training even without fully sampled data. We prove ShaRP minimizes an objective function involving a regularizer derived from the score functions of minimum mean square error (MMSE) restoration operators, and theoretically analyze its convergence. Empirically, ShaRP achieves state-of-the-art performance on tasks such as magnetic resonance imaging reconstruction and single-image super-resolution,  surpassing both denoiser- and diffusion-model-based methods without requiring retraining.
\end{abstract}

\section{Introduction}

Many problems in computational imaging, biomedical imaging, and computer vision can be viewed as \emph{inverse problems}, where the goal is to recover an unknown image from its noisy and incomplete measurements. Inverse problems are typically ill-posed, thus requiring additional prior information for accurate image reconstruction. While many approaches have been proposed for implementing image priors, the current research focuses on methods based on deep learning (DL)~\citep{McCann.etal2017, Ongie.etal2020, Kamilov.etal2023, Wen.etal2023}.

Deep neural networks trained as image denoisers are widely-used for specifying image priors for solving \emph{general} inverse problems~\citep{Romano.etal2017, Kadkhodaie.Simoncelli2021, Zhang.etal2022}. The combination of pre-trained Gaussian denoisers with measurement models has been shown to be effective in many inverse problems, including image super-resolution, deblurring, and medical imaging~\citep{Metzler.etal2018, Zhang.etal2017a, Meinhardt.etal2017, Dong.etal2019, Zhang.etal2019, Wei.etal2020, Zhang.etal2022} (see also the recent reviews
~\citep{Ahmad.etal2020, Kamilov.etal2023}). This success has led to active research on novel methods based on denoiser priors, their theoretical analyses, statistical interpretations, as well as connections to related approaches such as score matching and diffusion models~\citep{Venkatakrishnan.etal2013, Chan.etal2016, Romano.etal2017, Buzzard.etal2017, Reehorst.Schniter2019, Sun.etal2018a, Sun.etal2019b, Ryu.etal2019, Xu.etal2020, Liu.etal2021b, Cohen.etal2021a, Hurault.etal2022, hurault2022proximal, Laumont.etal2022, Gan.etal2023a, renaud2024plug}.

The mathematical relationship between denoising and the score function (the gradient of the log of the image distribution), known as the Tweedie's formula~\citep{robbins1956empirical,efron2011tweedie} seemingly implies that Gaussian denoising alone might be sufficient for learning priors, independent of the specific characteristics of an inverse problem. However, there is limited research exploring whether broader classes of priors based on pre-trained restoration models could outperform those based on Gaussian denoisers. In this paper, we present evidence that priors derived from deep models pre-trained as general restoration operators can surpass those trained exclusively for Gaussian denoising. We introduce a novel method called \emph{\textbf{S}toc\textbf{ha}stic deep \textbf{R}estoration \textbf{P}riors (ShaRP)}, which provides a principled approach to integrate an ensemble of general restoration models as priors to regularize inverse problems. ShaRP is conceptually related to several recent papers exploring priors specified using other types of image restoration operators, such as, for example, image super-resolution models~\citep{Zhang.etal2019, Liu.etal2020, Gilton.etal2021a, hurestoration}. However, unlike these methods, ShaRP provides a richer and more flexible representation of image priors by considering restoration models trained on a range of degradation types (e.g., various undersampling masks in MRI or blur kernels in image deblurring). By using more versatile restoration models, ShaRP improves upon traditional methods using Gaussian denoiser priors in two key ways: (a) ShaRP improved performance by using restoration models better suited to mitigating the structured artifacts that arise during inference, which are often linked to the characteristics of the underlying inverse problem. (b) Unlike Gaussian denoisers, the restoration models in ShaRP can often be directly trained in a self-supervised manner without fully sampled measurement data.

We present new theoretical and numerical results highlighting the potential of using an ensemble of restoration models as image priors.  Our first theoretical result introduces a novel notion of regularization for inverse problems corresponding to the average of likelihoods associated with the degraded observations of an image. The proposed regularizer has an intuitive interpretation as promoting solutions whose multiple degraded observations resemble realistic degraded images. We show that ShaRP seeks to minimize an objective function containing this regularizer. Our second theoretical result analyzes the convergence of ShaRP iterations when using both exact and inexact minimum mean squared error (MMSE) restoration operators. Numerically, we show the practical relevance of ShaRP by applying it to MRI reconstruction with varying undersampling patterns and rates, using a fixed-rate pre-trained MRI reconstruction network as a prior. We also show that ShaRP can use a pre-trained image deblurring model to perform single image super-resolution (SISR). Our numerical experiments show that ShaRP effectively adapts the pre-trained restoration model as a prior, outperforming existing methods based on image denoisers and diffusion models, and achieving state-of-the-art results. Our experiments additionally highlight the benefit of using restoration models as priors by considering a setting where only undersampled and noisy MRI data is available for pre-training the prior. In such cases, self-supervised training of a restoration model is feasible, whereas training a Gaussian denoiser requires fully sampled data.

\section{Background}

\textbf{Inverse Problems.} Many computational imaging tasks can be formulated as inverse problems, where the goal is to reconstruct an unknown image  $\xbm\in\R^n$ from its corrupted measurement 
\begin{equation}
\label{Eq:InverseProblem}
\ybm=\Abm\xbm + \ebm, 
\end{equation}
where $\Abm\in\R^{m\times n}$ is a measurement operator and $\ebm\in\R^m$ is the noise. A common approach to addressing inverse problems is to formulate them as an optimization problem 
\begin{equation}
    \label{equ:optimization}
    \xbmhat \in \argmin_{\xbm\in\R^n} f(\xbm) \quad\text{with}\quad f(\xbm) = g(\xbm) + h(\xbm)\ ,
\end{equation}
where $g$ is the data-fidelity term that quantifies the fit to the measurement $\ybm$ and $h$ is a regularizer that incorporates prior information on $\xbm$. For instance, typical functions used in imaging inverse problems are the least-squares term  $g(\xbm)=\frac{1}{2}\norm{\Abm\xbm-\ybm}_2^2$ and the total variation (TV) regularizer $h(\xbm)=\tau\norm{\Dbm\xbm}_1$, where $\Dbm$ is the image gradient and $\tau > 0$ is a regularization parameter. 

\textbf{Deep Learning.} DL has emerged as a powerful tool for addressing inverse problems~\citep{McCann.etal2017, Ongie.etal2020, Wen.etal2023}. Instead of explicitly defining a regularizer, DL methods use
deep neural networks (DNNs) to map the measurements to the desired images~\citep{Wang2016.etal, DJin.etal2017, Kang.etal2017, Chen.etal2017, delbracio2021projected, indi2023}. Model-based DL (MBDL) is a widely-used sub-family of DL algorithms that integrate physical measurement models with priors specified using CNNs (see reviews by~\cite{Ongie.etal2020, Monga.etal2021}). The
literature of MBDL is vast, but some well-known examples include plug-and-play priors (PnP), regularization by denoising (RED), deep unfolding (DU), compressed sensing using generative models (CSGM), and deep equilibrium models (DEQ)~\citep{Bora.etal2017, Romano.etal2017, zhang2018ista, Hauptmann.etal2018, Gilton.etal2021, Liu.etal2022a, hu2024spicer}. These approaches come with different trade-offs in terms of imaging performance, computational and memory complexity, flexibility, need for supervision, and theoretical understanding.

\textbf{Denoisers as Priors.} Score-based models (SBMs) are a powerful subset of DL methods for solving inverse problems that use deep Gaussian denoisers as imaging priors. Plug-and-Play (PnP) methods can be viewed as SBMs that incorporate denoisers within iterative optimization algorithms (see recent reviews~\citep{Ahmad.etal2020, Kamilov.etal2023}). These approaches construct a cost function by combining an explicit likelihood with a score function implicitly defined by the denoiser prior. Over the past few years, numerous variants of PnP have been developed~\citep{Venkatakrishnan.etal2013, Romano.etal2017, Metzler.etal2018, Zhang.etal2017a, Meinhardt.etal2017, Dong.etal2019, Zhang.etal2019, Wei.etal2020, Hurault.etal2022}, which has motivated an extensive research  into their theoretical properties and empirical effectiveness~\citep{Chan.etal2016, Buzzard.etal2017, Ryu.etal2019, Sun.etal2018a, Tirer.Giryes2019, Teodoro.etal2019, Xu.etal2020, Sun.etal2021, Cohen.etal2020, hu2022monotonically, Laumont.etal2022, hurault2022proximal, Gan.etal2023a, Cohen.etal2021a, fang2023s,renaud2024plug, hu2024a, renaud2024plugandplay, terris2024equivariant}.
 Diffusion Models (DMs) represent another category of SBMs; they are trained to learn the score function of the underlying probability distribution governed by stochastic differential equations (SDEs)~\citep{ho2020denoising,song2021scorebased}. Once trained, these models can be used as powerful priors for inverse problems by leveraging their learned score functions. Specifically, pre-trained DMs facilitate posterior sampling by guiding the denoising process to generate data consistent with observed measurements. This approach enables DMs to address inverse problems, often achieving impressive perceptual performance even for highly ill-posed inverse problems~\citep{chung2023diffusion, zhu2023denoising, wang2023zeroshot, feng2023score, sun2024provable, wu2024principled, song2024solving, hu2024learning, alccalar2024zero, zhao2024cosign, rout2024beyond, bai2024blind}.

\textbf{Restoration Networks as Priors.} In addition to denoiser-based methods, recent work has also considered using restoration models as implicit priors for solving inverse problems~\citep{Zhang.etal2019, Liu.etal2020, Gilton.etal2021a, hurestoration}. It has been observed that pre-trained restoration models can be effective priors for addressing unseen inverse problems, sometimes surpassing traditional denoiser-based approaches~\citep{hurestoration}. However, existing methods present two main limitations.
First, restoration models considered so far have relied on a fixed prior tailored to a specific degradation. Although deep restoration models can be trained in various settings---such as different blur kernels for image deblurring or diverse undersampling masks for MRI reconstruction---current approaches do not leverage this capability, limiting their robustness to diverse artifacts. Second, prior work has not explored the potential of learning restoration priors directly from undersampled measurements, without access to fully sampled data. Unlike Gaussian denoisers, training without fully sampled data is natural for restoration models~\citep{Yaman.etal2020, Liu.etal2020, Tachella.etal2022, Chen.etal2022, millard2023theoretical, gan2023self}. It is also worth highlighting the related work that has explored using corrupt measurements for training Ambient DMs~\citep{daras2023ambient, aali2024ambient}. Ambient DMs seek to sample from $p_\xbm$ using DMs trained directly on undersampled measurements. Thus, during inference Ambient DMs assume access to the image prior $p_\xbm$, while ShaRP only assumes access to the ensemble of likelihoods of multiple degraded observations.

\textbf{Our contribution.} \textbf{(1)} We propose ShaRP, a new framework for solving inverse problems leveraging a set of priors implicit in a pre-trained deep restoration network. ShaRP generalizes Regularization by Denoising (RED)~\citep{Romano.etal2017} and Stochastic Denoising Regularization (SNORE)~\citep{renaud2024plug} by using more flexible restoration operators and extends Deep Restoration Priors (DRP)~\citep{hurestoration} by using multiple restoration priors instead of relying on a single one. \textbf{(2)} We introduce a novel regularization concept for inverse problems that encourages solutions that produce degraded versions closely resembling real degraded images. For example, our regularizer favors an MR image solution only if its various degraded versions are consistent with the characteristics of actual degraded MR images. \textbf{(3)} We show that ShaRP minimizes a composite objective that incorporates our proposed regularizer. We provide a theoretical analysis of its convergence for both exact and approximate MMSE restoration operators. \textbf{(4)} We implement ShaRP with both supervised and self-supervised restoration models as priors and test it on two inverse problems: compressed sensing MRI (CS-MRI) and single-image super-resolution (SISR). Our results highlight the capability of restoration models to achieve state-of-the-art performance. Notably, in the MRI context, we show that restoration networks trained directly on subsampled and noisy MRI data can serve as effective priors, a scenario where training traditional Gaussian denoisers is infeasible.

\begin{figure}[t] 
  \centering
  \includegraphics[width=.855\textwidth]{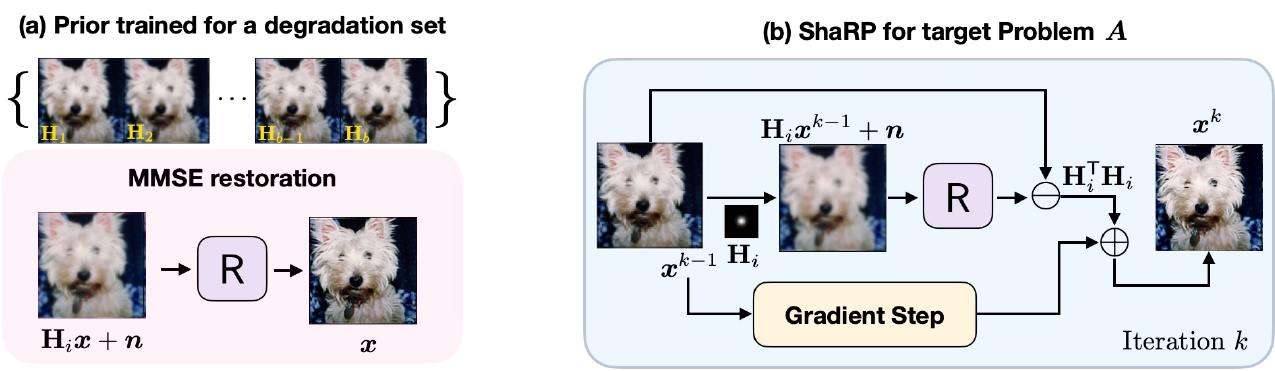}
  \caption{A restoration network trained on a set of tasks $\{\Hbf_i\}$ can be used as a prior within ShaRP to address different target inverse problems without the need for retraining.
} 
  \label{fig:method}
\end{figure}

\section{Stochastic Deep Restoration Priors}

ShaRP is presented in Algorithm~\ref{alg:ShaRP}. It considers a prior based on a deep restoration model $\Rsf(\sbm, \Hbf)$ pre-trained using the family of $b$ degradation operators $\{\Hbf_1, \cdots, \Hbf_b\}$, such as blur kernels or MRI masks. More specifically, the deep restoration model $\Rsf$ is trained to solve the following set of restoration problems
\begin{equation}
\label{Eq:RestorationProblem}
\sbm_i = \Hbf_i\xbm + \nbm_i \quad\text{with}\quad \xbm \sim p_\xbm, \quad \nbm_i \sim \Ncal(0, \sigma^2 \Ibf), \quad i \in \{1, \cdots, b\},
\end{equation}
where $\nbm_i$ are the AWGN vectors with variance $\sigma^2$ and $p_\xbm$ denotes the probability distribution of the target images. Importantly, the restoration problems~\eqref{Eq:RestorationProblem} are used exclusively for training $\Rsf$ and do not need to match the target inverse problem~\eqref{Eq:InverseProblem}, which involves the measurement operator $\Abm$. 

\begin{algorithm}[t]
\caption{Stochastic deep Restoration Priors (ShaRP)}
\label{alg:ShaRP}
\begin{algorithmic}[1]
\State \textbf{input: } Initial value $\xbm^0 \in \R^n$, $\gamma > 0$, $\sigma > 0$, and $\tau > 0$
\For{$k = 1, 2, 3, \dots$}
\State $\text{Select a degradation operator: }\; \Hbf \in \{\Hbf_1, \cdots, \Hbf_b\}$
\State $\sbm \leftarrow \Hbf\xbm^{k-1} + \nbm$ with $\nbm \sim \Ncal(0, \sigma^2\Ibf)$
\State $\xbm^k \leftarrow \xbm^{k-1} - \gamma \hat{\nabla} f(\xbm^{k-1})$
\Statex $\quad\quad\quad$ where $\hat{\nabla}f(\xbm^{k-1}) \defn \nabla g (\xbm^{k-1}) + (\tau/\sigma^2) \Hbf^{\Tsf}\Hbf (\xbm^{k-1} - \Rsf(\sbm, \Hbf))$
\EndFor
\end{algorithmic}
\end{algorithm}

Our analysis below shows that $\nablahat f$ corresponds to a stochastic approximation of an objective function of form $f = g + h$. Similar to traditional stochastic gradient methods, ShaRP can be implemented using various selection strategies for the degradation operators. Although the algorithm above is described using a finite set of $b$ operators, our theoretical analysis adopts a more general approach by considering the degradation operator $\Hbf$ to be sampled from a distribution $\Hbf \sim p_{\Hbf}$.

Each iteration of ShaRP has an intuitive interpretation, where the next solution is obtained by combining the gradient of the data-fidelity term $\nabla g$ and the residual of restored image corresponding to the selected degradation operator. When $\Hbf_i = \Ibf$ for all $i \in \{1, \cdots, b\}$, then the restoration prior reduces to the Gaussian denoiser, and ShaRP can be viewed as an instance of the Regularization by Denoising (RED) method~\citep{Romano.etal2017} and Stochastic Denoising Regularization (SNORE)~\citep{renaud2024plug}. On the other hand, when $b = 1$, then ShaRP can be viewed as the instance of the Deep Restoration Priors (DRP) method~\citep{hurestoration}. Thus, ShaRP can be viewed as a more versatile generalization of both that can account for various degradation operators.



\section{Theoretical analysis of ShaRP}
\label{Sec:TheoreticalAnalysis}

We present two theoretical results on ShaRP. The first establishes the regularizer minimized by ShaRP, while the second analyzes its convergence with inexact MMSE operators.

Consider a restoration model that perform MMSE estimation of $\xbm \in \R^n$ for problems~\eqref{Eq:RestorationProblem}
\begin{equation}
\label{Eq:MMSERestorator}
\Rsf^\ast(\sbm, \Hbf) = \E\left[\xbm | \sbm, \Hbf \right] = \int \xbm \, p(\xbm | \sbm, \Hbf) \d \xbm = \frac{1}{p(\sbm| \Hbf)} \int \xbm \, G_\sigma(\sbm-\Hbf\xbm){p_{\xbm}}(\xbm) \d \xbm.
\end{equation}
where we used the probability density of the observation $\sbm$ conditioned on the operator $\Hbf$ 
\begin{equation}
\label{Eq:ObservationPDF}
p(\sbm | \Hbf) = \int G_\sigma(\sbm - \Hbf\xbm) p_\xbm(\xbm) \d \xbm.
\end{equation}
The function $G_\sigma$ in~\eqref{Eq:ObservationPDF} denotes the Gaussian density function with the standard deviation $\sigma > 0$.

\medskip\noindent
We define the ShaRP regularizer as
\begin{equation}
\label{Eq:ExpReg}
 h(\xbm)= \tau \E_{\sbm \sim G_\sigma(\sbm - \Hbf\xbm), \Hbf \sim p_\Hbf} \left[ -\log \, p(\sbm|\Hbf)\right],
\end{equation}
where $\tau > 0$ is the regularization parameter and $p_{\Hbf}$ is the distribution of all considered degradation operators. The regularizer $h$ is minimized when the degraded versions of $\xbm$ have a high likelihood under the probability density of degraded observations, $p(\sbm | \Hbf)$. In other words, $h$ identifies an image as a valid solution if its degraded observations resemble realistic degraded images.

\medskip\noindent
We are now ready to state our first theoretical result.
\begin{theorem}
\label{Thm:sgd}
Assume that the prior density $p_\xbm$ is non-degenerate over $\R^n$ and let $\Rsf^\ast$ be the MMSE restoration operator~\eqref{Eq:MMSERestorator} corresponding to the restoration problems~\eqref{Eq:RestorationProblem}. 
Then, we have that
\begin{equation} 
\label{Eq:ShaRPFullGrad}
\nabla h(\xbm) = \frac{\tau}{\sigma^2} \left(\E_{\sbm \sim G_\sigma(\sbm - \Hbf\xbm), \Hbf \sim p_\Hbf}\left[\Hbf^\Tsf\Hbf(\xbm-\Rsf^\ast(\sbm, \Hbf))\right]\right),
\end{equation}
where $h$ is the ShaRP regularizer in~\eqref{Eq:ExpReg}.
\end{theorem}
The proof is provided in the appendix. Note that the expression within the square parenthesis in~\eqref{Eq:ShaRPFullGrad} matches the ShaRP update in Line 4 of Algorithm~\ref{alg:ShaRP}, which directly implies that ShaRP using the exact MMSE restoration operator $\Rsf^\ast$ is a stochastic gradient method for minimizing $f = g + h$, where $g$ is the data-fidelity term and $h$ is the ShaRP regularizer in~\eqref{Eq:ExpReg}.

We now present the convergence analysis of ShaRP under a restoration operator $\Rsf$ that \emph{approximates} the true MMSE restoration operator $\Rsf^\ast$. For a given degraded observation $\sbm = \Hbf\xbm+\nbm$ with $\Hbf \sim p_{\Hbf}$ and $\nbm \sim \Ncal(0, \sigma^2\Ibf)$, we define the stochastic gradient used by ShaRP
\begin{equation}
\nablahat f(\xbm) = \nabla g(\xbm) + \nablahat h(\xbm) \quad\text{with}\quad \nablahat h(\xbm) \defn \frac{\tau}{\sigma^2}\Hbf^\Tsf\Hbf(\xbm-\Rsf(\sbm, \Hbf)).
\end{equation}
Since $\Rsf$ is an inexact MMSE restoration operator, we also define the bias vector
\begin{equation}
\label{Eq:BiasTerm}
\bbm(\xbm) = \frac{\tau}{\sigma^2}\E_{\sbm \sim G_\sigma(\sbm - \Hbf\xbm), \Hbf \sim p_\Hbf}\left[\Hbf^\Tsf\Hbf(\Rsf^\ast(\sbm, \Hbf)-\Rsf(\sbm, \Hbf))\right],
\end{equation}
which quantifies the average difference between the exact and inexact MMSE restoration operators.
Our analysis requires three assumptions that jointly serve as sufficient conditions for our theorem.
\begin{assumption}
\label{As:LipschitzFunction}
The function $f$ has a finite minimum $f^\ast > -\infty$ and the gradient $\nabla f$ is Lipschitz continuous with constant $L > 0$.
\end{assumption}
This is a standard assumption used in the analysis of gradient-based algorithms (see~\citep{Nesterov2004}, for example). It is satisfied by a large number of functions, including the traditional least-squares data-fidelity function.
\begin{assumption}
\label{As:sgd}
The stochastic gradient has a bounded variance for all $\xbm \in \R^n$, which means that there exists a constant $\nu > 0$ such that
    \[
\E \left[\left\| \nablahat f(\xbm) - \E\left[\nablahat f(\xbm)\right]\right\|_2^2\right] \leq \nu^2,
    \]
where expectations are with respect to $\Hbf \sim p_{\Hbf}$ and $\sbm \sim G_\sigma(\sbm-\Hbf\xbm)$.
\end{assumption}
This is another standard assumption extensively used in the analysis of online or stochastic optimization algorithms~\citep{Bertsekas2011, Ghadimi.Lan2016}. 
\begin{assumption}
\label{As:bias}
The bias is bounded, which means that there exists $\varepsilon > 0$ such that for all $\xbm \in \R^n$
    \[
 \|\bbm(\xbm)\|_2 \leq \varepsilon.
    \]
\end{assumption}
Note that our only assumption on the bias is that it is bounded, which is a relatively mild assumption in the context of biased stochastic gradient methods~\citep{Demidovich.etal2023}.

\begin{theorem}
\label{Thm:inexactmmse}
Run ShaRP for $t \geq 1$ iterations using the step-size $0 < \gamma \leq 1/L$ under Assumptions~\ref{As:LipschitzFunction}-\ref{As:bias}. Then, the sequence $\xbm^k$ generated by ShaRP satisfies
\[
\E \left[\frac{1}{t} \sum_{k = 1}^t \|\nabla f(\xbm^{k-1})\|_2^2\right] \leq \frac{2}{\gamma t} (f(\xbm^0) - f^\ast) + \gamma L\nu^2 + \varepsilon^2.
\]
\end{theorem}
The proof is provided in the appendix. This theorem shows that \emph{in expectation}, ShaRP minimizes the norm of the gradient $\nabla f$ up to an error term that has two components, $\gamma L \nu^2$ and $\epsilon^2$. Since the first component depends on $\gamma$, it can be made as small as desired by controlling the step-size $\gamma$. The second component only depends on the magnitude of the bias $\varepsilon$, which, in turn, directly depends on the accuracy of the restoration operator relative to the true MMSE restoration operator $\Rsf^\ast$.

\section{Numerical Results}
We numerically validate ShaRP on two inverse problems of the form $\ybm = \Abm\xbm + \ebm$: (\emph{Compressive Sensing MRI (CS-MRI)} and (b) \emph{Single Image Super Resolution (SISR)}. In both cases, $\ebm$ represents additive white Gaussian noise (AWGN). For the data-fidelity term in eq.~\eqref{equ:optimization}, we use the $\ell_2$-norm loss for both problems. Quantitative performance is evaluated using Peak Signal-to-Noise Ratio (PSNR) and Structural Similarity Index (SSIM). Additionally, for the SISR task, we include the Learned Perceptual Image Patch Similarity (LPIPS) metric to evaluate perceptual quality. Additional numerical results are provided in the supplementary material.

\begin{figure}[t]
  \centering
  \includegraphics[width=.995\textwidth]{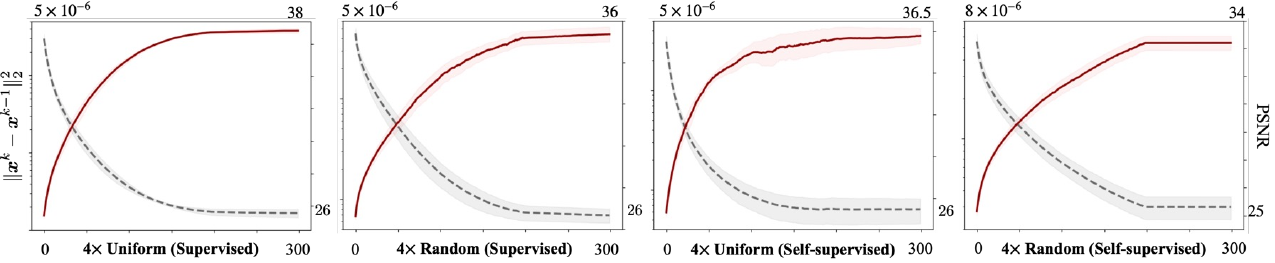}
  \caption{Convergence of ShaRP for $4\times$ accelerated MRI reconstruction on the fastMRI dataset. \textbf{(a)-(b)} depict the convergence behavior of ShaRP using restoration operators trained in a supervised manner, while \textbf{(c)-(d)} correspond to those trained in a self-supervised manner. The plots illustrate the average distance $\Vert\xbm^{k} - \xbm^{k-1} \Vert^2_2$ and PSNR relative to the ground truth, as a function of the iteration number, with shaded regions representing the standard deviation. Note the stable convergence of ShaRP with both types of priors.
}
 \label{fig:convergence}

\end{figure}

\subsection{CS-MRI setting}
The  measurement of CS-PMRI can be modeled as $\ybm = \Pbm\Fbm\Sbm\xbm + \ebm$, where $\Pbm$ is the k-space subsampling pattern, $\Fbm$ is the Fourier transform operator, $\Sbm = ( \Sbm_1,\cdots,\Sbm_{n_c})$ are the multi-coil sensitivity maps, and $\ebm$ is the noise vector.

\textbf{Dataset.} We simulated multi-coil subsampled measurements using T2-weighted human brain MRI data from the open-access fastMRI dataset, which comprises 4,912 fully sampled multi-coil slices for training and 470 slices for testing. Each slice has been cropped into a complex-valued image with dimensions $320 \times 320$. The coil sensitivity maps for each slice are precomputed using the ESPIRiT algorithm~\citep{Uecker.etal2014}. We simulated a Cartesian sampling pattern that subsamples along the $k_y$ dimension while fully sampling along the $k_x$ dimension.

\textbf{Ensemble of Restoration Priors for CS-MRI.} Recent methods, such as InDI~\citep{indi2023} and I$^2$SB~\citep{liu2023i2sb},  introduce controllable processes for training an ensemble of restoration priors, where each prior functions as an MMSE restoration operator tailored to a specific setting. Building on this approach, we trained an $8\times$ uniform subsampling CS-MRI model with 8 distinct masks as our restoration prior. Similar to InDI, we decompose the original MRI degradation operator $\Mbm$ into several convex combinations of the original task $\Mbm$ and the identity mapping $\Ibf$, represented by the new degradation operator $\Hbf_\alpha = (1-\alpha)\Ibf + \alpha\Mbm$, with $\alpha$ controlling the degradation level. By selecting a range of $\alpha$ values, we create an ensemble of restoration tasks. Training the restoration network $\Rsf$ to handle all these tasks allows it to function as an ensemble of MMSE restoration operators, $\Rsf(\sbm, \Hbf_\alpha) = \E\left[\xbm | \sbm, \Hbf_\alpha \right]$. We used the MSE loss to train the restoration model.

\textbf{Training restoration priors without groundtruth.}
When fully-sampled ground truth images are not available for training restoration priors, MRI restoration priors can be trained in self-supervised manner~\citep{Yaman.etal2020, millard2023theoretical, gan2023self, hu2024spicer}. In self-supervised training, rather than using the ground-truth image as the label, a separate subsampled measurement serves as the label. In such cases, we can train our priors using a weighted $\ell_2$
  loss function, following the self-supervised approach in~\citep{gan2023self}. We thus train the $8\times$ uniform subsampling CS-MRI model to handle eight distinct restoration operators, each corresponding to a different sampling mask.
 
Additional details on supervised and self-supervised restoration network training and our CS-MRI sampling masks are in Section~\ref{Sup:Sec:MRI_experiment} of the appendix.

With the pre-trained $8\times$ models as ensembles of restoration priors, we evaluate ShaRP's performance across a variety of configurations, including two sub-sampling rates (4$\times$ and 6$\times$), two mask types (uniform and random), and three noise levels ($\sigma = 0.005$, $0.01$, and $0.015$).

\begin{table}[h]\small
\centering
\renewcommand\arraystretch{1.1}
\setlength{\tabcolsep}{1.8pt}
\begin{tabular}{ccccccccccccc}
\toprule
      & \multicolumn{6}{c}{4$\times$ Uniform}                                                        & \multicolumn{6}{c}{6$\times$ Uniform}                                                                                \\ \hline
Noise level & \multicolumn{2}{c}{$\sigma$ = 0.005} & \multicolumn{2}{c}{$\sigma$ = 0.010} & \multicolumn{2}{c}{$\sigma$ = 0.015} & \multicolumn{2}{c}{$\sigma$ = 0.005} & \multicolumn{2}{c}{$\sigma$ = 0.010} & \multicolumn{2}{c}{$\sigma$ = 0.015} \\ \hline
Metrics     & PSNR                   & SSIM                 & PSNR                    & SSIM                    & PSNR                    & SSIM                    & PSNR                    & SSIM & PSNR                    & SSIM & PSNR                    & SSIM                   \\ \hline   
Zero-filled          &26.93                         &0.848                                                 &26.92                         &0.847                                         &26.90                      &0.848           &22.62               &0.728     &22.60                         &0.726                         &22.59                         &0.721                         \\
TV          &31.17                         &0.923                                                 &31.08                         &0.921                   &30.91                         &0.915                      &25.00                         &0.806     &24.94                         &0.803                         &24.33                         &0.755                         \\

PnP-FISTA          &35.88                         &0.938                                                 &31.14                         &0.894                   &30.32                         &0.846                      &26.30                         &0.822     &25.97                         &0.786                         &25.46                         &0.747                         \\
PnP-ADMM         &\secondbest{35.76}                         &\secondbest{0.941}                                                 &32.36                         &0.878                   &30.66                         &0.838                      &26.13                         &0.808     &25.90                         &0.776                         &25.51                         &0.742                         \\
DRP           & 35.52                         & 0.936   & 32.32                         & 0.914                      & 30.57                         & 0.901                         & 29.51                         & 0.872                        & 28.52                         & 0.882                        & 28.35                         & 0.876                        \\

DPS              &32.62                         &0.888                                                 &31.39                         &0.870                   &30.29                         &0.856                      &30.53                         &0.862     &29.41                         &0.843                         &28.63                         &0.830                         \\
DDS                     &35.21                         &0.937                                                 &\secondbest{35.03}                         &\secondbest{0.935}                   &\secondbest{34.51}                         &\secondbest{0.925}                      &\secondbest{31.02}                         &\secondbest{0.889}     &\secondbest{30.84}                         &\secondbest{0.888}                         &\secondbest{30.79}                         &\secondbest{0.888}                         \\

\midrule
ShaRP           & \best{37.59}                         &\best{0.963}  &\best{35.81} & \best{0.951} & \best{34.92}                         &\best{0.942} & \best{33.42}                         &\best{0.940} & \best{32.86}                         &\best{0.932} & \best{32.09}                         &\best{0.922}                          \\
\bottomrule   
\end{tabular}


\caption{Quantitative comparison of ShaRP with several baselines for CS-MRI using uniform masks at undersampling rates of 4 and 6 on fastMRI dataset. The \textbf{\color[HTML]{D52815}best} and \underline{\color[HTML]{00008A}second best} results are highlighted. Notably, ShaRP outperforms SOTA methods based on denoisers and diffusion models.
}
\label{tab:mri_uniform}

\end{table}

\textbf{Baselines.}  ShaRP was compared against several baseline methods, including denoiser-based approaches (PnP-FISTA~\citep{Kamilov.etal2017}, PnP-ADMM~\citep{Chan.etal2016}) and diffusion model-based methods (DPS~\citep{chung2023diffusion}, DDS~\citep{chungdecomposed}). To highlight the advantages of using a stochastic set of restoration operators, we also compared ShaRP with the DRP method~\citep{hurestoration}, which applies only a single restoration operator. Additional details related to the baseline methods can be found in Section~\ref{Sup:Sec:MRI_experiment} of the appendix.

\begin{figure}[h]
  \centering
  \includegraphics[width=.985\textwidth]{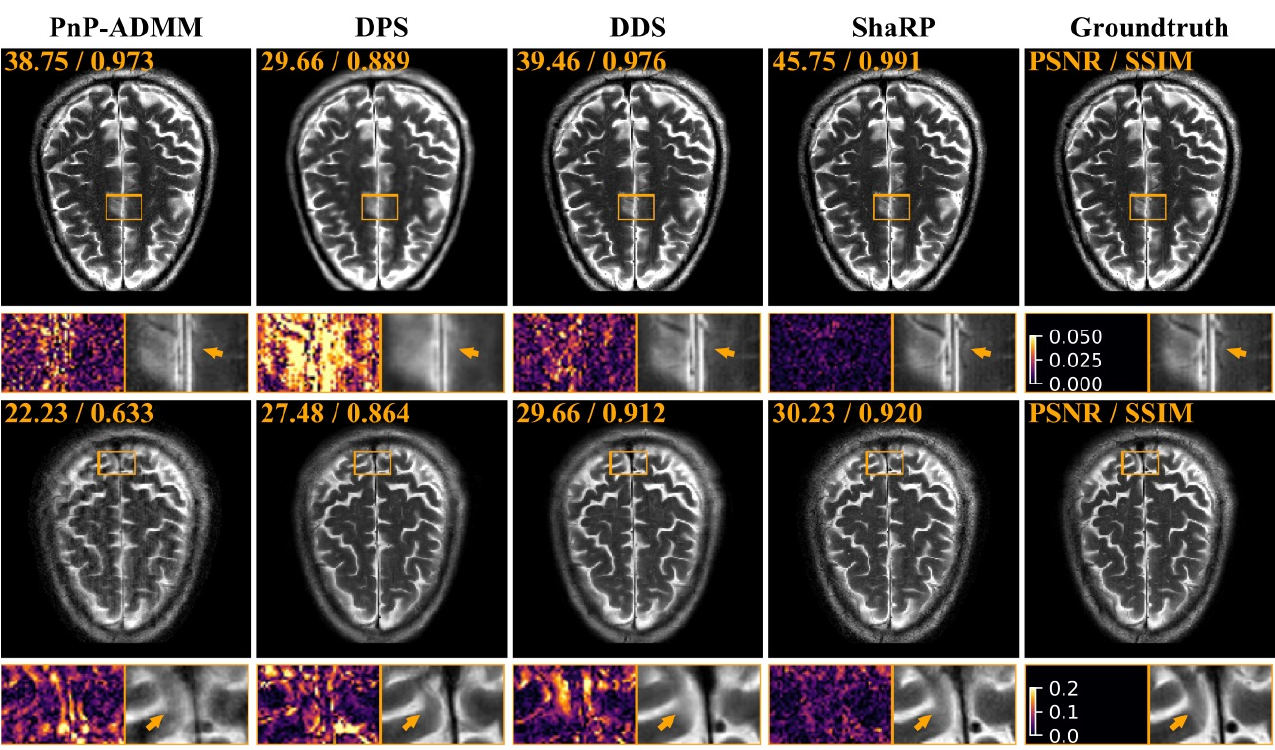}
  \caption{Visual comparison of ShaRP with baseline methods on CS-MRI. The top row shows results for a $4\times$ random mask with noise $\sigma = 0.005$, and the bottom row for a $6\times$ random mask with noise $\sigma = 0.015$. PSNR and SSIM values are in the top-left corner of each image. Error maps and zoomed-in areas highlight differences. Notably, ShaRP with stochastic priors outperforms state-of-the-art methods using denoiser and diffusion model priors.}
  \label{fig:mri_main}

\end{figure}

\textbf{Results with supervised MMSE Restoration operator.} Figure~\ref{fig:convergence} illustrates the convergence behavior of ShaRP on the test set with an acceleration factor of $R = 6$ and additional noise $\sigma = 0.01$. Table~\ref{tab:mri_uniform} provides a quantitative comparison of reconstruction performance across different acceleration factors and noise levels using a uniform sub-sampling mask. In all configurations, ShaRP consistently outperforms the baseline methods.  The use of a set of restoration operators clearly enhances ShaRP’s performance, highlighting the effectiveness of employing multiple operators to maximize the regularization information provided by the restoration model. Figure~\ref{fig:mri_main} presents visual reconstructions for two test scenarios, where ShaRP accurately recovers fine brain details, particularly in the zoomed-in regions, while baseline methods tend to oversmooth or introduce artifacts. These results highlight ShaRP's superior ability to manage structured artifacts and preserve fine details, outperforming both denoiser-based and diffusion model-based methods.

\textbf{Results with self-supervised MMSE Restoration operator.}
We further evaluate ShaRP's performance using an restoration model, learned in a self-supervised manner, as introduced in~\citep{gan2023self}. In this setting, we compare ShaRP against two classical methods for CS-MRI reconstruction without groundtruth: TV~\citep{block2007undersampled} and GRAPPA~\citep{Griswold2002} and a recent state-of-the-art self-supervised deep unrolling method: SPICER~\citep{hu2024spicer}. As shown in Table~\ref{tab:self_mri}, even in scenarios where only incomplete measurements ($8\times$ subsampled measurement) are available, ShaRP can effectively apply a self-supervised trained restoration prior to various reconstruction tasks. ShaRP using self-supervised restoration prior even outperforms DPIR and DPS that use Gaussian denoisers trained using fully-sampled ground truth images (see Table~\ref{tab:random_mri} in the appendix). Note that given only undersampled measurements, training Gaussian denoisers is not feasible.

\begin{table}[h]\small
\centering
\renewcommand\arraystretch{1.1}
\setlength{\tabcolsep}{1.8pt}

\definecolor{almond}{rgb}{0.94, 0.87, 0.8}
\newcolumntype{a}{>{\columncolor{almond}}c}

\begin{tabular}{ccccccccccccc}
\toprule
      & \multicolumn{6}{c}{4$\times$ Random}                                                        & \multicolumn{6}{c}{6$\times$ Random}                                                                                \\ \hline
Noise level & \multicolumn{2}{c}{$\sigma$ = 0.005} & \multicolumn{2}{c}{$\sigma$ = 0.010} & \multicolumn{2}{c}{$\sigma$ = 0.015} & \multicolumn{2}{c}{$\sigma$ = 0.005} & \multicolumn{2}{c}{$\sigma$ = 0.010} & \multicolumn{2}{c}{$\sigma$ = 0.015} \\ \hline
Metrics     & PSNR                   & SSIM                 & PSNR                    & SSIM                    & PSNR                    & SSIM                    & PSNR                    & SSIM & PSNR                    & SSIM & PSNR                    & SSIM                   \\ \hline

\rowcolor{almond}  PnP-ADMM                                                          &28.83                         &0.842                                                 &28.39                         &0.816                   &27.70                         &0.786                      &25.59                         &0.776     &25.19                         &0.740                         &24.93                         &0.728                         \\

ADMM-TV          &{28.14}                        &{0.866}                                                 &{28.06}                         &{0.863}                   &{27.96}                         &{0.859}                      &{24.55}                         &{0.782}     &{24.33}                         &{0.750}                         &{24.28}                         &{0.736}                        \\

GRAPPA          &28.09                         &0.792                                                 &25.39                         &0.699                    &23.94                         &0.649                      &{25.67}                         &{0.737}     &23.72                         &0.646                         &22.51                         &0.595                         \\

SPICER          &\secondbest{31.87}                         &\secondbest{0.901 }                                                &\secondbest{31.67}                         &\secondbest{0.889 }                   &\secondbest{31.50}                       &\secondbest{0.887 }                     &\secondbest{30.18}                         &\secondbest{0.871}     &\secondbest{30.05}                          &\secondbest{0.863}                          &\secondbest{30.01}                          &\secondbest{0.860}                          \\

\midrule
$\text{ShaRP}^\textbf{self}$           &\best{33.87}                         &\best{0.909}                                                 &\best{33.64}                         &\best{0.900}                  & \best{33.21}                         & \best{0.892}                 &\best{30.87}                         &\best{0.899}     &\best{30.36}                         &\best{0.890}                         &\best{30.21}                         &\best{0.875}                         \\

\bottomrule  
\end{tabular}
\vspace{1.5mm}
\caption{PSNR (dB) and SSIM values for ShaRP with a self-supervised pre-trained restoration operator, compared to several baselines for CS-MRI with random undersampling masks at rates of 4 and 6 on the fastMRI dataset. The \textbf{\color[HTML]{D52815}best} and \underline{\color[HTML]{00008A}second best} results are highlighted. For reference, the highlighted row presents a PnP method using a Gaussian denoiser, which requires fully sampled data for training. Note the excellent performance of ShaRP even using priors trained without fully-sampled ground-truth data.}
\label{tab:self_mri}
\end{table}

\begin{figure}[h]
  \centering
  \includegraphics[width=.985\textwidth]{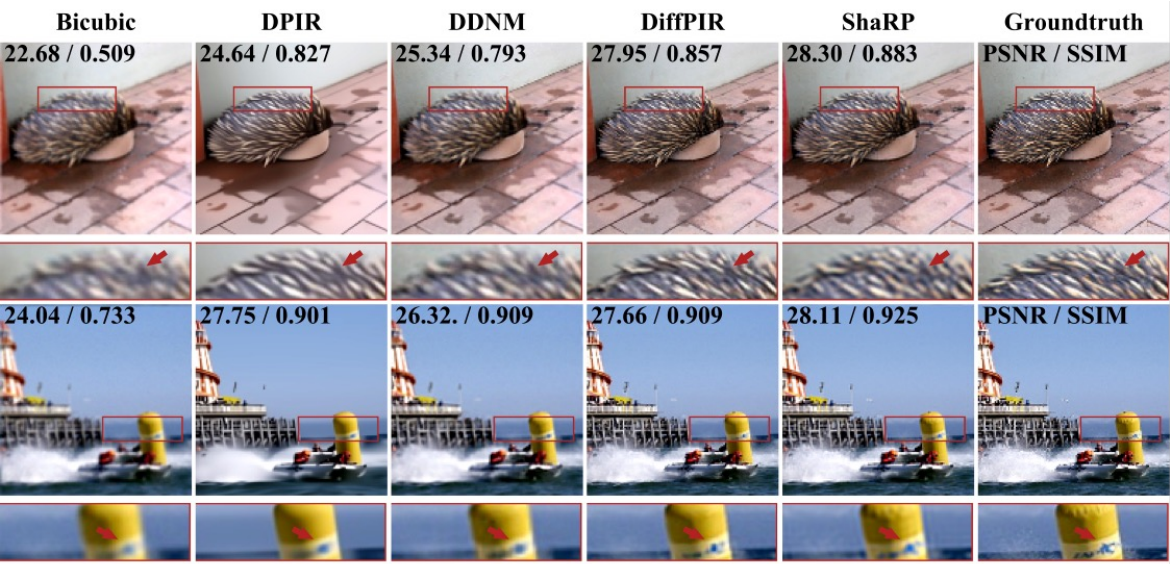}
  \caption{Visual comparison of ShaRP with several well-known methods on $2\times$ SISR. The top row shows results for SISR with gaussian blur kernel with $\sigma = 1.25$, while the bottom row shows results for SISR with gaussian blur kernel with $\sigma = 1.5$. The quantities in the top-left corner of each image provide PSNR and SSIM values for each method. The squares at the bottom of each image visualize the zoomed area in the image.}
 \label{fig:sisr_main}

\end{figure}

\subsection{Single Image Super Resolution (SISR)}
The measurement operator in SISR can be written as $\Abm = \Sbm \Kbm$, where $\Kbm$ represents convolution with the blur kernel, and $\Sbm$ performs standard $d$-fold down-sampling. In our experiments, we use two Gaussian blur kernels $\kbm$ , each with distinct standard deviations (1.25 and 1.5), and with down-sampling factor of 2. Both noisy and noise-free cases are considered to evaluate the noisy robustness of ShaRP. We randomly selected 100 images from the ImageNet test set, as provided in DiffPIR\footnote{\href{https://github.com/yuanzhi-zhu/DiffPIR/tree/main/testsets}{https://github.com/yuanzhi-zhu/DiffPIR/tree/main/testsets}}. 

\textbf{Ensemble of Restoration Priors for Image Deblurring.} Following the approach used to train our CS-MRI restoration prior, we decompose the original deblurring task represented by the Gaussian blur operator $\Kbm$ into convex combinations of the original task and the identity mapping $\Isf$. This results in a new degradation operator defined as $\Hbf_\alpha = (1 - \alpha)\Isf + \alpha\Kbm$, where $\alpha$ controls the degradation level. By varying $\alpha$, we generate multiple degradation operators, allowing us to train the restoration network $\Rsf$ to handle all these operators. This training enables $\Rsf$ to function as an ensemble of MMSE restoration operators, expressed as $\Rsf(\sbm, \Hbf_\alpha) = \E[\xbm \mid \sbm, \Hbf_\alpha]$, where $\sbm$ is the degraded image and $\xbm$ is the original image. The original deblurring operator $\Kbm$ corresponds to convolution with a Gaussian blur kernel of size $31 \times 31$ and standard deviation 3. More details on the deblurring restoration network training are in Section~\ref{Sup:Sec:SISR_experiment} of the Appendix.
 
 \textbf{Baselines.} We compared ShaRP with several baseline methods, including DPIR~\citep{Zhang.etal2022}, DPS~\citep{chung2023diffusion}, DDNM~\citep{wang2023zeroshot}, and DiffPIR~\citep{zhu2023denoising}. DPIR represents the state-of-the-art (SOTA) PnP method that uses pre-trained denoisers as priors to address SISR. In contrast, DPS, DDNM, and DiffPIR use different sampling strategies to leverage pre-trained unconditional diffusion models for solving SISR. More baseline details can be found in Section~\ref{Sup:Sec:SISR_experiment} of the Appendix.
 
  \textbf{Results on SISR with deblurring prior.} Figure~\ref{fig:sisr_main} shows the visual reconstruction results for two settings with different blur kernels. As demonstrated, ShaRP successfully recovers most features and maintains high data consistency with the available measurements. Table~\ref{tab:sisr_main} provides a quantitative comparison of ShaRP against other baseline methods, evaluated across various blur kernels and noise levels. ShaRP achieves the highest PSNR and SSIM values but ranks second in perceptual performance (LPIPS). This is consistent with the SOTA perceptual performance of DMs on SISR. However, note how the use of a deblurring prior within ShaRP enables it to recover fine details, ensuring overall competitiveness of the perceptual quality of the ShaRP solutions.

\begin{table}[h]\small
\centering
\renewcommand\arraystretch{1.1}
\setlength{\tabcolsep}{1.8pt}
\begin{tabular}{ccccccccccccc}
\toprule
      & \multicolumn{6}{c}{{\includegraphics[width=0.035\textwidth]{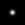}}}                                                      & \multicolumn{6}{c}{{\includegraphics[width=0.035\textwidth]{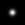}}     }                                                                           \\ \hline
Noise level & \multicolumn{3}{c}{Noiseless} & \multicolumn{3}{c}{$\sigma$ = 0.01} & \multicolumn{3}{c}{Noiseless} & \multicolumn{3}{c}{$\sigma$ = 0.01}  \\ \hline
Metrics     & PSNR                   & SSIM                 & LPIPS                    & PSNR                   & SSIM                 & LPIPS                    & PSNR                   & SSIM                 & LPIPS& PSNR                   & SSIM                 & LPIPS                   \\ \hline   
DPIR              &28.10                   &0.809                         &0.305                         &28.05                   &0.807                         &0.308                      &27.90                         &0.803     &0.314                         &27.87                         &0.800                         &0.314                         \\

DDNM           &27.53                         & 0.786                                                 &0.240                         &27.49                   & 0.784                         & 0.246                      &27.02                         & 0.764     & 0.264                         &27.01                         &0.763                         & 0.267                         \\

DPS           &24.68                         &0.661                                                 &0.395                         &24.60                   &0.657                         &0.399                      &24.50                         &0.657     &0.403                         &24.44                         &0.655                         &0.406                         \\

DiffPIR           &28.92                         &0.852                                                 &\best{0.152}                         &28.63                   &0.839                         &\best{0.169}                      &\secondbest{28.59}                         &\secondbest{0.834}     &\best{0.172}                         &\secondbest{28.02}                         &\secondbest{0.819}                         &\best{0.185}                         \\

DRP           &\secondbest{29.28}                         &\secondbest{0.868}                                                 &0.207                         &\secondbest{28.87}
&\secondbest{0.848}                         &0.248                      &{28.24}                         &{0.836}     &0.235                         &{28.01}                         &{0.822}                         &0.278                         \\

ShaRP           &\best{30.09}                         &\best{0.891}                                                 &\secondbest{0.179}                         &\best{29.03}                   &\best{0.852}                         &\secondbest{0.223}                      &\best{29.28}                         &\best{0.872}     &\secondbest{0.209}                         &\best{28.06}                         &\best{0.821}                         &\secondbest{0.268}                         \\

\bottomrule  
\end{tabular}
\caption{Quantitative comparison of ShaRP with several baselines for SISR based on two different blur kernels on ImageNet dataset. The \textbf{\color[HTML]{D52815}best} and \underline{\color[HTML]{00008A}second best} results are highlighted. Notably, ShaRP outperforms SOTA methods based on denoisers and diffusion models.
}
\label{tab:sisr_main}

\end{table}

\section{Conclusion}
The work presented in this paper proposes a new ShaRP method for solving imaging inverse problems by using pre-trained restoration network as a prior, presents its theoretical analysis, and applies the method to two well-known inverse problems. Unlike previous approaches that relied on Gaussian denoisers or a single restoration prior, our method uses a set of restoration priors, each corresponding to different degradation settings. The numerical validation shows that ShaRP benefits from stochastically using multiple degradation priors, leading to better results. A key conclusion is the potential effectiveness of exploring priors beyond those defined by traditional Gaussian denoisers and a specific restoration operator.

\section*{Ethics Statement}
To the best of our knowledge this work does not give rise to any significant ethical concerns.

\section*{Acknowledgments}
The authors extend their sincere gratitude to Regev Cohen and Ashok Popat for their valuable feedback. This research was supported by NSF CCF-2043134. We also thank Google for their generous contribution.

\bibliography{iclr2025_conference}
\bibliographystyle{iclr2025_conference}

\newpage
\appendix

\section{Theoretical Analysis of ShaRP}

\subsection{Proof of Theorem~\ref{Thm:sgd}}
\label{Sup:Sec:Theorem1}

\begin{theorem*}
Assume that the prior density $p_\xbm$ is non-degenerate over $\R^n$ and let $\Rsf^\ast$ be the MMSE restoration operator~\eqref{Eq:MMSERestorator} corresponding to the restoration problems~\eqref{Eq:RestorationProblem}. 
Then, we have that
\begin{equation*} 
\nabla h(\xbm) = \frac{\tau}{\sigma^2} \left(\E_{\sbm \sim G_\sigma(\sbm - \Hbf\xbm), \Hbf \sim p_\Hbf}\left[\Hbf^\Tsf\Hbf(\xbm-\Rsf^\ast(\sbm, \Hbf))\right]\right),
\end{equation*}
where $h$ is the ShaRP regularizer in~\eqref{Eq:ExpReg}.
\end{theorem*}

\begin{proof}
The ShaRP regularizer \( h(\xbm) \) is defined as
\begin{align}
 \nonumber h(\xbm) &= \tau \E_{\sbm \sim G_\sigma(\sbm - \Hbf\xbm), \Hbf \sim p_\Hbf} \left[ -\log p(\sbm|\Hbf)\right] \\
 & = - \tau \int p(\Hbf) \left[ \int G_\sigma(\sbm - \Hbf\xbm) \log p(\sbm|\Hbf) \d\sbm \right] \d\Hbf, \label{Eq:ShaRPRegExp}
\end{align}
where \( G_\sigma \) is the Gaussian probability density with variance \( \sigma^2 \) and \( p(\sbm|\Hbf) \) is the likelihood function for the degraded observation given the operator \( \Hbf \). The expectation over \( p(\Hbf) \) accounts for the randomness of the restoration operator \( \Hbf \).

We start by relating the MMSE restoration operator to the score of the degraded observation
\begin{equation*}
\nabla p(\sbm | \Hbf) = \frac{1}{\sigma^2} \int \left(\Hbf\xbm-\sbm\right) G_\sigma(\sbm-\Hbf\xbm)p_\xbm(\xbm) \d \xbm,
\end{equation*}
where $p_\xbm$ is the prior. By using the definition of the MMSE estimator, we obtain the relationship
\begin{equation}
\label{Eq:GeneralizedTweedie}
\nabla \log p(\sbm | \Hbf) = \frac{1}{\sigma^2} \left(\Hbf \Rsf^\ast(\sbm, \Hbf)-\sbm\right).
\end{equation}

Consider the function inside the parenthesis in the expression for the ShaRP regularizer~\eqref{Eq:ShaRPRegExp}
\begin{equation*}
 \rho (\zbm) \defn (G_\sigma \ast \log p_{\sbm | \Hbf})(\zbm) = \int G_\sigma(\zbm-\sbm) \, \log p(\sbm | \Hbf) \d \sbm,
\end{equation*}
where $\zbm$ has the same dimensions as $\sbm$ and $\ast$ denotes convolution. The gradient of $\rho$ is given by
\begin{align*}
 \nabla \rho(\zbm) &= (\nabla G_\sigma * \log p_{\sbm|\Hbf})(\zbm) = ( G_\sigma * \nabla\log p_{\sbm|\Hbf})(\zbm)\\
 &= \frac{1}{\sigma^2} \int G_\sigma(\zbm - \sbm) \left[\Hbf \Rsf^\ast(\sbm, \Hbf)-\sbm\right] \d \sbm \\
 &= \frac{1}{\sigma^2} \left(\Hbf \int \Rsf^\ast(\sbm, \Hbf) G_\sigma(\zbm-\sbm) \d \sbm - \zbm \right)
\end{align*}
where we used~\eqref{Eq:GeneralizedTweedie}. By using $\zbm = \Hbf\xbm$, we write the gradient with respect to $\xbm$
\begin{equation*}
\nabla_\xbm \rho(\Hbf\xbm) = \frac{1}{\sigma^2} \Hbf^\Tsf\Hbf \left( \int \Rsf^\ast(\sbm, \Hbf)G_\sigma (\sbm-\Hbf\xbm) \d \sbm - \xbm\right)
\end{equation*}
By using this expression in~\eqref{Eq:ShaRPRegExp}, we obtain the desired result
\begin{align*}
\nabla h(\xbm) 
&= -\frac{\tau}{\sigma^2} \left[\int p(\Hbf) \int G_\sigma(\sbm-\Hbf\xbm) \left(\Hbf^\Tsf\Hbf(\Rsf^\ast(\sbm, \Hbf)-\xbm)\right) \d \sbm \d \Hbf\right] \\
&= \frac{\tau}{\sigma^2} \E_{\sbm \sim G_\sigma(\sbm-\Hbf\xbm), \Hbf\sim p_\Hbf} \left[\Hbf^\Tsf\Hbf(\xbm - \Rsf^\ast(\sbm, \Hbf))\right].
\end{align*}
\end{proof}

\subsection{Proof of Theorem 2}
\label{Sup:Sec:ConvergenceProof}

\begin{theorem*}
Run ShaRP for $t \geq 1$ iterations using the step-size $0 < \gamma \leq 1/L$ under Assumptions~\ref{As:LipschitzFunction}-\ref{As:bias}. Then, the sequence $\xbm^k$ generated by ShaRP satisfies
\[
\E \left[\frac{1}{t} \sum_{k = 1}^t \|\nabla f(\xbm^{k-1})\|_2^2\right] \leq \frac{2}{t} (f(\xbm^0) - f^\ast) + \gamma L\nu^2 + \varepsilon^2.
\]
\end{theorem*}

\begin{proof}

First note that from the definition of the bias in eq.~\eqref{Eq:BiasTerm}, we have that
\begin{equation}
\label{Eq:BiasedStochGrad}
\E \left[\nablahat f(\xbm^{k-1}) \,|\, \xbm^{k-1}\right] = \nabla f(\xbm^{k-1}) + \bbm(\xbm^{k-1}),
\end{equation}
where the expectation is with respect to $\sbm \sim G_\sigma(\sbm-\Hbf\xbm^{k-1})$ and $\Hbf \sim p_{\Hbf}$. In order to simplify the notation, we will drop these subscripts from the expectations in the analysis below.

Consider the iteration $k \geq 1$ of ShaRP with inexact MMSE operator
\begin{align*}
f(\xbm^k) &\leq f(\xbm^{k-1}) + \nabla f(\xbm^{k-1})^\Tsf(\xbm^k - \xbm^{k-1}) + \frac{L}{2}\|\xbm^k-\xbm^{k-1}\|_2^2 \\
 &= f(\xbm^{k-1}) -\gamma  \nabla f(\xbm^{k-1})^\Tsf\nablahat f(\xbm^{k-1}) + \frac{L\gamma^2}{2}\| \nablahat f(\xbm^{k-1}) \|^2,
\end{align*}
where in the first line we used the Lipschitz continuity of $\nabla f$. By taking the expectation with respect to $\sbm \sim G_\sigma(\sbm-\Hbf\xbm^{k-1})$ and $\Hbf \sim p_{\Hbf}$ on both sides of this expression, we get
\begin{align*}
\E[f(\xbm^k) | \xbm^{k-1}] &\leq f(\xbm^{k-1}) - \gamma \nabla f(\xbm^{k-1})^\Tsf(\nabla f(\xbm^{k-1})+\bbm(\xbm^{k-1})) + \frac{L\gamma^2}{2} \E\left[\|\nablahat f(\xbm^{k-1})\|_2^2 | \xbm^{k-1}\right] \\
&\leq f(\xbm^{k-1}) - \frac{\gamma}{2}\|\nabla f(\xbm^{k-1})\|_2^2 + \frac{\gamma}{2}\|\bbm(\xbm^{k-1})\|_2^2 \\
&\quad\quad+ \frac{L\gamma^2}{2}\left(\E\left[\|\nablahat f(\xbm^{k-1})\|_2^2 | \xbm^{k-1} \right] - \left(\E[\nablahat f(\xbm^{k-1}) | \xbm^{k-1}]\right)^2\right)\\
&\leq f(\xbm^{k-1}) - \frac{\gamma}{2}\|\nabla f(\xbm^{k-1})\|_2^2 + \frac{\gamma \varepsilon^2}{2} + \frac{L\gamma^2 \nu^2}{2}.
\end{align*}

In the second row, we completed the square, applied eq.~\eqref{Eq:BiasedStochGrad}, and used the assumption that $\gamma \leq 1/L$. In the third row, we used the variance and bias bounds in Assumptions~\ref{As:sgd} and~\ref{As:bias}. By rearranging the expression, we get the following bound
\begin{equation*}
\|\nabla f(\xbm^{k-1})\|_2^2 \leq \frac{2}{\gamma}\left(f(\xbm^{k-1})-\E[f(\xbm^k) | \xbm^{k-1}]\right) + L\gamma\nu^2 + \varepsilon^2
\end{equation*}
By taking the total expectation, averaging over $t$ iterations, and using the lower bound $f^\ast$, we get the desired result
\begin{equation*}
    \E\left[\frac{1}{t}\sum_{k = 1}^t \|\nabla f(\xbm^{k-1})\|_2^2\right] \leq \frac{2}{\gamma t}(f(\xbm^0)-f^\ast) + L\gamma\nu^2 + \varepsilon^2.
\end{equation*}

\end{proof}

\newpage

\section{Experiment Details}
\label{Sup:Sec:Experiment_settings}
\subsection{Implementation details of CS-MRI tasks}
 \label{Sup:Sec:MRI_experiment}

\textbf{Subsampling pattern for CS-MRI.}
In this paper, we explored two types of subsampling patterns for MRI reconstruction tasks. All undersampling masks were generated by first including a set number of \emph{auto-calibration signal (ACS)} lines, ensuring a fully-sampled central k-space region.

\begin{figure}[h]
  \centering
  \includegraphics[width=.985\textwidth]{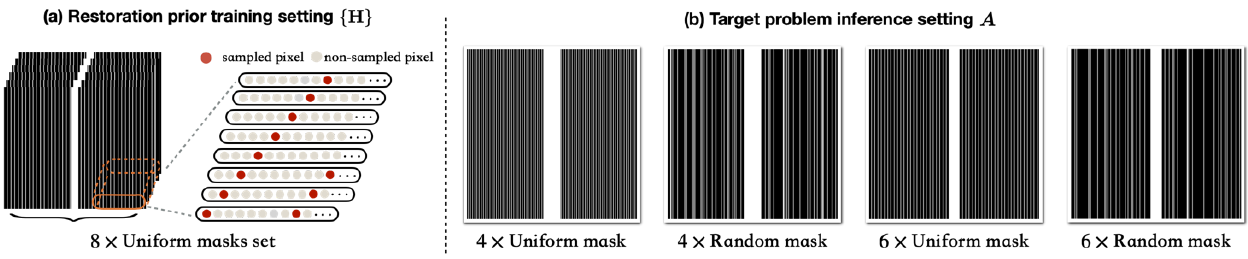}
  \caption{Illustration of the undersampling masks used for CS-MRI in this work. \textbf{(a)} The eight different $8\times$ uniform masks used for training the restoration prior. \textbf{(b)} The inference setting for ShaRP, demonstrating how the prior trained on the masks in \textbf{(a)} can be applied to solve other problems without retraining.}
\label{fig:mri_mask}
\end{figure}

Figure~\ref{fig:mri_mask} illustrates the k-space trajectories for both random and uniform (equidistant) subsampling at acceleration factors of 4, 6, and 8. Notably, different patterns were used for training and testing. During training, our restoration prior was only trained on a uniform mask with a subsampling rate of 6. However, for inference, we employed both uniform and random masks at subsampling rates of 4 and 6, creating a mismatch between the pre-trained restoration prior and the test configurations.

\begin{algorithm}[h]
  \caption{Supervised Training of CS-MRI Restoration Network} 
  \label{alg:supervised_training_mri}
  \begin{algorithmic}
     \Require  \text{dataset}: $p(\xbm)$, \text{sampling operator set:} $\{\Mbm_1,\Mbm_1,\cdots,\Mbm_1\}$, \text{Restoration model}: $\Rsf_\theta(\cdot, \alpha)$ 
    \State \textbf{repeat:} 
    \State \quad $\xbm \sim p(\xbm)$, $\Mbm \sim \{\Mbm_1,\Mbm_2,\cdots,\Mbm_8\}$, $\ebm \sim \Ncal(0, \sigma^2\Ibf)$, $\alpha \sim \mathcal{U}([0,1]) $
    \State \quad $\ybm = \Mbm\xbm + \ebm$
    \State \quad  $\min_\theta \left\| \Rsf_\theta \left((1-\alpha) \xbm + \alpha \Mbm^\Tsf\ybm ; \alpha \right) - \xbm \right\|^2_2$
    \State \textbf{until converge}
  \end{algorithmic}
  \end{algorithm}

\subsubsection{Implementation of Supervised Prior for CS-MRI}

\textbf{Models training for supervised case.}
We use the same U-Net architecture as employed in the official implementation of DDS\footnote{\url{https://github.com/HJ-harry/DDS}} for $\Rsf(\cdot; \alpha)$. For the supervised learning case, we select 1,000 different $\alpha$ values to train the model, following the $\alpha$ schedule outlined by I$^2$SB~\citep{liu2023i2sb}. The model is trained with Adam optimizer with a learning rate of $5 \times 10^{-5}$. As shown in Algorithm~\ref{alg:supervised_training_mri}, we train our supervised learning model using eight different masks for $8\times$ uniform sampling CS-MRI reconstruction. In the pseudocode, $\{\Mbm_1,\Mbm_2,\cdots,\Mbm_8\}$ represent the eight different MRI degradation operators, each defined by a unique sampling pattern, as shown in Figure~\ref{fig:mri_mask} (a). This results in a total of 8,000 possible combinations of $\alpha$ values and sampling masks, effectively creating an ensemble of restoration priors during training.

\textbf{Inference with a Subset of the Ensemble (Supervised Case).} During inference, to simplify computation and focus on the most effective priors, we use only a subset of the supervised trained ensemble. Specifically, we fix the $\alpha$ value to a particular choice (e.g., $\alpha = 0.5$) and use the 8 different sampling masks $\{\Mbm_1,\Mbm_2,\cdots,\Mbm_8\}$, resulting in 8 restoration priors.

\textbf{Step size and regularization parameter.} To ensure fairness, for each problem setting, each method—both proposed and baseline—is fine-tuned for optimal PSNR using 10 slices from a validation set separate from the test set. The same step size $\gamma$ and regularization parameter $\tau$ are then applied consistently across the entire test set.

\textbf{Baseline details.}  We compare ShaRP with several variants of denoiser- and diffusion model-based methods. For denoiser-based approaches, we include PnP-FISTA~\citep{Kamilov.etal2023}, PnP-ADMM~\citep{Chan.etal2016}. PnP-FISTA and PnP-ADMM correspond to the FISTA and ADMM variants of PnP, both utilizing AWGN denoisers built on DRUNet~\citep{Zhang.etal2022}. For diffusion model-based methods, we compare with DPS~\citep{chung2023diffusion} and DDS~\citep{chungdecomposed}, which use pre-trained diffusion models as priors and apply different posterior sampling strategies to address general inverse problems. We use the same pre-trained diffusion model configuration as outlined in the DDS paper. For all baseline methods, we fine-tuned their parameters to maximize the PSNR value. Notably, both the DRUNet denoiser and the diffusion model were trained using the same dataset employed for training our restoration prior. For a fair comparison, the diffusion model pre-trained for DDS and DPS use the same network architecture as our restoration network . All models are trained from scratch on the fastMRI training set, following the architecture settings provided in DDS\footnote{\href{https://github.com/HJ-harry/DDS}{https://github.com/HJ-harry/DDS}}. We also compared with method that also use the deep restoration prior to solve general inverse problem: DRP~\citep{hurestoration}. For DRP, we utilize the same pre-trained restoration network as in ShaRP. However, instead of employing a set of degradation priors, DRP uses a single fixed prior. For a fair comparison, we selected the optimal fixed prior—defined by a fixed $\alpha$ and subsampling mask—based on PSNR performance on the validation set, and applied it accordingly.

\subsubsection{Implementation of Self-Supervised Prior for CS-MRI}

\begin{algorithm}[h]
  \caption{Self-Supervised Training of CS-MRI Restoration Network} 
  \label{alg:self_supervised_training_mri}
  \begin{algorithmic}
     \Require  \text{dataset:} $p(\ybm_i, \Mbm_i, \ybm_j, \Mbm_j)$, \text{Restoration model}: $\Rsf_\theta(\cdot)$ 
    \State \textbf{repeat:} 
        \State \quad $\ybm_i, \Mbm_i, \ybm_j, \Mbm_j \sim p(\ybm_i, \Mbm_i, \ybm_j, \Mbm_j)$, $\ebm \sim \Ncal(0, \sigma^2\Ibf)$
    \State \quad $\min_\theta \left\| \Mbm_j \Rsf_\theta \left(\Mbm_i^\Tsf (\ybm_i + \ebm) \right) - \ybm_j \right\|^2_\Wbm$
    \State \textbf{until converge}
  \end{algorithmic}
  \end{algorithm}

\textbf{Models training for (Self-Supervised Case).}
For self-supervised training, the ground truth reference $\xbm$ is not available as a label. Instead, as shown in Algorithm~\ref{alg:self_supervised_training_mri}, we work with pairs of subsampled measurements, $y_i$ and $y_j$, along with their corresponding sampling operators, $\Mbm_i$ and $\Mbm_j$. These paired measurements exhibit significant overlap within the shared \emph{auto-calibration signal (ACS)} region, which increases the weighting of these overlapping k-space regions. Following the approach proposed by SSDEQ~\citep{gan2023self}, we introduce a diagonal weighting matrix $\Wbm$ to account for the oversampled regions in the loss function. By incorporating this weighted loss, we are able to train our MMSE restoration operator using incomplete measurements alone. Furthermore, unlike the supervised case where we use the combination of $\alpha$ values to form an ensemble, in the self-supervised setting, we construct the ensemble using only eight different sampling masks across the entire dataset. 

\textbf{Inference Using All Restoration Priors (Self-Supervised Case).} During inference in the self-supervised setting, we utilize all 8 restoration priors corresponding to the different sampling masks. By incorporating the entire ensemble, we fully leverage its capacity to remove the artifacts and enhance reconstruction performance.

\textbf{Step size and regularization parameter.} To ensure fairness, for each problem setting, each method—both proposed and baseline—is fine-tuned for optimal PSNR using 10 slices from a validation set separate from the test set. The same step size $\gamma$ and regularization parameter $\tau$ are then applied consistently across the entire test set.

\textbf{Baseline details.}  In the self-supervised setting, we compared ShaRP with two widely used traditional methods: TV~\citep{block2007undersampled} and GRAPPA~\citep{Griswold2002}, both of which address the restoration problem without requiring fully-sampled references. Additionally, we included SPICER~\citep{hu2024spicer}, a recent state-of-the-art self-supervised deep unrolling method designed for MRI reconstruction using only pairs of undersampled measurements. To ensure consistency, we trained the SPICER model on the same amount of paired data used for training our restoration prior in the $8\times$ uniform CS-MRI setting and applied it to other CS-MRI configurations.

\clearpage
\subsection{Implementation details of SISR tasks}
\label{Sup:Sec:SISR_experiment}
\begin{algorithm}[h]
  \caption{Gaussian Deblurring Restoration  network training} 
  \label{alg:supervised_training_sisr}
  \begin{algorithmic}
     \Require  \text{dataset:}$p(\xbm, \ybm)$, \text{Gaussian blur operator: $\Kbm$},  $\Rsf_\theta(\cdot, \alpha)$ 
    \State \textbf{repeat:} 
    \State \quad $\xbm \sim p(\xbm)$, $\ebm \sim \Ncal(0, \sigma^2\Ibf)$, $\alpha \sim \mathcal{U}([0,1]) $
    \State \quad $\min_\theta \left\| \Rsf_\theta \left((1-\alpha) \xbm + \alpha\Kbm\xbm  ; \alpha \right) - \xbm \right\|^2_2$
    \State \textbf{until converge}
  \end{algorithmic}
  \end{algorithm}

\textbf{Restoration Model training.}
We use the same U-Net architecture as the Gaussian deblurring model provided by I$^2$SB\footnote{\href{https://github.com/NVlabs/I2SB}{https://github.com/NVlabs/I2SB}}. Utilizing the pre-trained checkpoints from their repository, we fine-tune our model accordingly. Specifically, we align with their codebase and configure the model type to OT-ODE to satisfy our MMSE restoration operator assumption.

To create an ensemble of restoration priors, we consider a family of degradation operators that are convex combinations of the identity mapping $\Ibf$ and the Gaussian blur operator $\Kbm$. The blurring operator $\Kbm$ corresponds to convolution with a Gaussian blur kernel of size $31 \times 31$ and standard deviation 3.  Specifically, we define the degradation operator as $\Hbf_\alpha = (1 - \alpha)\Isf + \alpha\Kbm$, where $\alpha \in [0, 1]$ controls the degradation level. By varying $\alpha$, we generate multiple degradation operators, allowing us to train the restoration network $\Rsf$ to handle all these operators, expressed as $\Rsf(\sbm, \Hbf_\alpha) = \E\left[\xbm | \sbm, \Hbf_\alpha \right]$, where $\sbm$ is the degraded image and $\xbm$ is the original image. 

We select 1,000 different $\alpha$ values from the interval $[0, 1]$, following the $\alpha$ schedule outlined by I$^2$SB~\citep{liu2023i2sb}. This results in 1,000 different degradation operators $\Hbf_\alpha$, effectively creating an ensemble of restoration priors during training. The model is trained using the Adam optimizer with a learning rate of $5 \times 10^{-5}$.

\textbf{Inference with a Subset of the Ensemble.} During inference, to simplify computation and focus on the most effective priors, we use only a subset of the supervised trained ensemble. Specifically, we select 6 $\alpha$ values, resulting in 6 restoration priors. 

\textbf{Step size and regularization parameter.} To ensure fairness, for each problem setting, each method—both proposed and baseline—is fine-tuned for optimal PSNR using 5 images from a validation set separate from the test set. The same step size $\gamma$ and regularization parameter $\tau$ are then applied consistently across the entire test set.

\textbf{Baseline details.} We compare ShaRP against several denoiser- and diffusion model-based methods. For denoiser-based approaches, we evaluate DPIR~\citep{Zhang.etal2022}, which relies on half-quadratic splitting (HQS) iterations with DRUNet denoisers. For diffusion model-based methods, we compare with DPS~\citep{chung2023diffusion}, DDNM~\citep{wang2023zeroshot}, and DiffPIR~\citep{zhu2023denoising}. These methods all use the same pre-trained diffusion models as priors, but each employs a distinct posterior sampling strategy to solve general inverse problems. We specifically use the pre-trained diffusion model from DiffPIR. We also compared with method that also use the deep restoration prior to solve general inverse problem: DRP~\citep{hurestoration}. For DRP, we utilize the same pre-trained deblurring network as in ShaRP. However, instead of employing a set of degradation priors, DRP uses a single fixed prior. For a fair comparison, we selected the optimal fixed prior—defined by a fixed $\alpha$ based on PSNR performance on the validation set, and applied it accordingly. For all baselines, we fine-tuned their parameters to maximize PSNR performance. Notably, the diffusion model backbone for all diffusion-based baselines was trained on the same dataset used to train our restoration prior.

\newpage
\section{Additional results for CS-MRI}
\label{Sup:Sec:discussion_mri}

\subsection{Performance of ShaRP for random subsampling setting}
Due to space constraints, we present only the quantitative performance for the uniform subsampling setting in the main paper. In this section, we further evaluate ShaRP's performance on random subsampling setting, with two sub-sampling rates (4$\times$ and 6$\times$), and three noise levels ($\sigma = 0.005$, $0.01$, and $0.015$).

\begin{figure}[h]
  \centering
  \includegraphics[width=.895\textwidth]{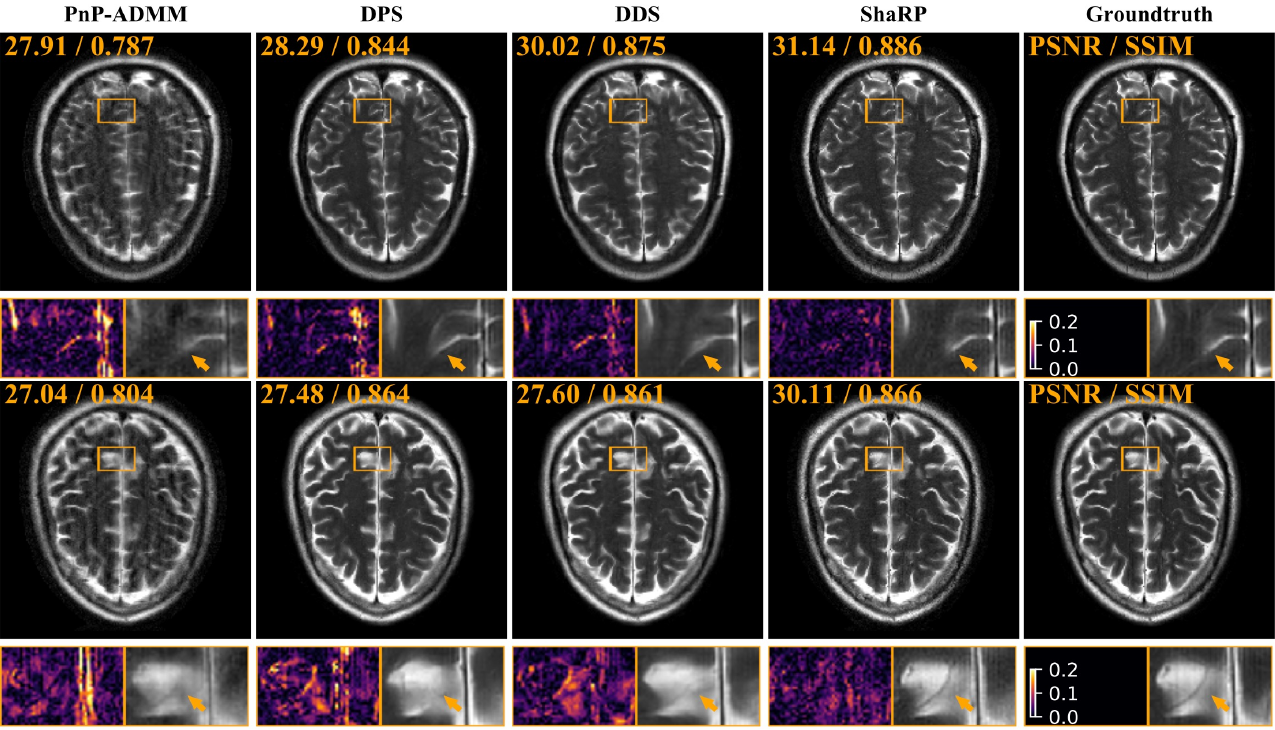}
  \caption{Visual comparison of ShaRP with baseline methods on CS-MRI for $6\times$ random sampling mask with noise $\sigma = 0.015$. PSNR and SSIM values are in the top-left corner of each image. Error maps and zoomed-in areas highlight differences. Notably, ShaRP with stochastic priors outperforms state-of-the-art methods using denoiser and diffusion model priors.}
  \label{Sup:fig:mri_random_6x}

\end{figure}

Table~\ref{tab:random_mri} provides a quantitative comparison of reconstruction performance across different acceleration factors and noise levels using a uniform sub-sampling mask. In all configurations, ShaRP consistently outperforms the baseline methods.  The use of a set of restoration operators clearly enhances ShaRP’s performance, highlighting the effectiveness of employing multiple operators to maximize the regularization information provided by the restoration model. Figure~\ref{Sup:fig:mri_random_6x} presents visual reconstructions for two test scenarios, where ShaRP accurately recovers fine brain details, particularly in the zoomed-in regions, while baseline methods tend to oversmooth or introduce artifacts. These results highlight ShaRP's superior ability to manage structured artifacts and preserve fine details, outperforming both denoiser-based and diffusion model-based methods.

\begin{table}[h]\small

\centering
\renewcommand\arraystretch{1.1}
\setlength{\tabcolsep}{1.8pt}
\begin{tabular}{ccccccccccccc}
\toprule
      & \multicolumn{6}{c}{4$\times$ Random}                                                        & \multicolumn{6}{c}{6$\times$ Random}                                                                                \\ \hline
Noise level & \multicolumn{2}{c}{$\sigma$ = 0.005} & \multicolumn{2}{c}{$\sigma$ = 0.010} & \multicolumn{2}{c}{$\sigma$ = 0.015} & \multicolumn{2}{c}{$\sigma$ = 0.005} & \multicolumn{2}{c}{$\sigma$ = 0.010} & \multicolumn{2}{c}{$\sigma$ = 0.015} \\ \hline
Metrics     & PSNR                   & SSIM                 & PSNR                    & SSIM                    & PSNR                    & SSIM                    & PSNR                    & SSIM & PSNR                    & SSIM & PSNR                    & SSIM                   \\ \hline   
Zero-filled          &25.83                         &0.815                   &25.81                         &0.812                   &25.76                         &0.807                      &22.68                         &0.724     &22.67                         &0.722                         &22.67                         &0.719                         \\
TV          &28.14                         &0.866                                                 &28.06                         &0.863                   &27.96                         &0.859                      &24.55                         &0.782     &24.33                         &0.750                         &24.28                         &0.736                        \\
PnP-FISTA                                                           &29.31                         &0.863                                                 &28.40                         &0.817                   &27.49                         &0.799                      &26.01                         &0.797     &25.63                         &0.756                         &24.94                         &0.717                        \\
PnP-ADMM                                                          &28.83                         &0.842                                                 &28.39                         &0.816                   &27.70                         &0.786                      &25.59                         &0.776     &25.19                         &0.740                         &24.93                         &0.728                         \\

DRP           & 29.97                         & 0.880   & 29.37                         & 0.839                      & 28.31                         & 0.794                         & 26.98                         & 0.866                        & 26.78                          & 0.853                        & 26.49                         & 0.821                        \\

DPS          &31.72                     &0.874  &30.45                     &0.857                                                     &29.50                     &0.843                   & 30.32                         &0.856                         &29.36                         &0.824                         &27.99                         &0.810                         \\
DDS     &\secondbest{32.41}                     &\secondbest{0.910}                          &\secondbest{32.37}                     &\secondbest{0.906}  &\secondbest{32.25}  &\secondbest{0.901}                     &\secondbest{30.59}                     &\secondbest{0.876}   & \secondbest{30.35}                         &\secondbest{0.874}                         &\secondbest{30.31}                         & \secondbest{0.879}                        \\ \midrule

ShaRP            & \best{34.66}                         &\best{0.949}                                                 & \best{33.57}                    & \best{0.920}                   &\best{33.18}  &\best{0.931}    & \best{31.53}                   & \best{0.924}         & \best{31.46}                        & \best{0.918}                        & \best{31.45}                   & \best{0.914}                   \\
\bottomrule  
\end{tabular}
\caption{Quantitative comparison of ShaRP with several baselines for CS-MRI using random masks at undersampling rates of 4 and 6 on fastMRI dataset. The \textbf{\color[HTML]{D52815}best} and \underline{\color[HTML]{00008A}second best} results are highlighted. Notably, ShaRP outperforms SOTA methods based on denoisers and diffusion models.
}
\label{tab:random_mri}
\end{table}

\clearpage
\subsection{Performance of additional baseline methods on matched and mismatched settings}
\begin{figure}[h]
  \centering
  \includegraphics[width=.795\textwidth]{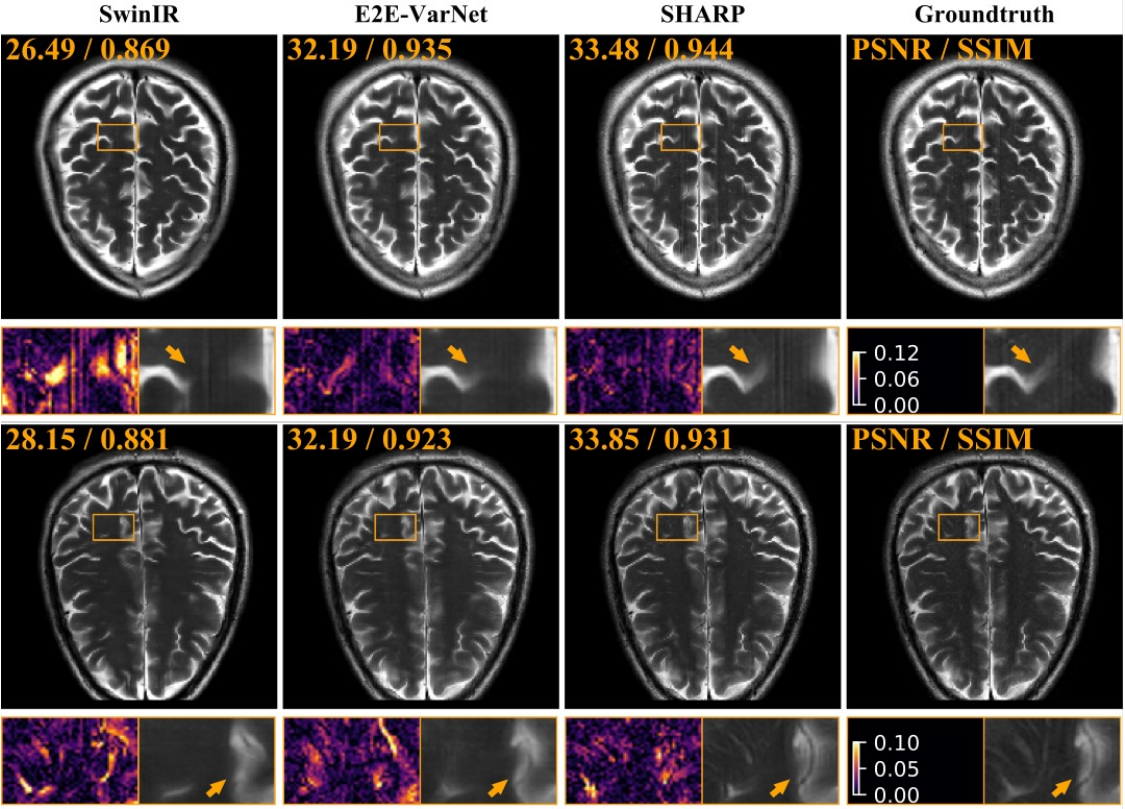}
  \caption{Visual comparison of ShaRP with task-specific baseline methods on CS-MRI for $6\times$ random sampling mask with noise $\sigma = 0.015$. PSNR and SSIM values are in the top-left corner of each image. Error maps and zoomed-in areas highlight differences. Notably, ShaRP with stochastic priors outperforms state-of-the-art methods using denoiser and diffusion model priors.}
  \label{Sup:fig:mismatched}

\end{figure}

In this section, we highlight an important observation: pre-trained restoration networks typically exhibit poor generalization to mismatched settings. We chose two commonly used methods (SwinIR~\citep{liang2021swinir} and E2E-VarNet~\citep{Sriram.etal2020}) for the specific setting of CS-MRI. We trained them on the same $8\times$ uniform subsampling setting as our restoration prior and directly applied them to solve both matched and mismatched problems, as ShaRP did. As shown in the Table~\ref{tab:mismatch_mri}, the baseline method's performance drops significantly under mismatched conditions, whereas ShaRP maintains stable performance and convergence guarantees. This demonstrates ShaRP's ability to adapt pre-trained restoration models as priors and use it to solve problems under mismatched settings. As shown in the Figure~\ref{Sup:fig:mismatched}, due to the mismatched settings, the two baseline methods suffer from over-smoothing, lack important details, and exhibit artifacts, whereas ShaRP still provides high-quality reconstruction performance. This indicates that ShaRP can balance data fidelity and the artifact removal capabilities of the prior model, leading to an artifact-free reconstruction that preserves important details.

\begin{table}[h]\small
\centering
\renewcommand\arraystretch{1.1}
\setlength{\tabcolsep}{1.8pt}

\definecolor{almond}{rgb}{0.94, 0.87, 0.8}
\newcolumntype{a}{>{\columncolor{almond}}c}

\begin{tabular}{ccccccccccc}
\toprule
Settings & \multicolumn{2}{c}{$4\times$ Uniform} & \multicolumn{2}{c}{$4\times$ Random} & \multicolumn{2}{c}{$6\times$ Uniform} & \multicolumn{2}{c}{$6\times$ Random} &\multicolumn{2}{a}{$8\times$ Uniform} \\ \midrule

Metrics     & PSNR                   & SSIM                 & PSNR                    & SSIM                    & PSNR                    & SSIM                    & PSNR                    & SSIM  &\cellcolor{almond}PSNR                    & \cellcolor{almond}SSIM                 \\ \hline   


SwinIR           & 24.78                         & 0.849   & 25.09                         & 0.841                                   & 29.55                         & 0.907                   & 27.98                         & 0.819 &\cellcolor{almond}29.37                         & \cellcolor{almond}0.898                           \\
E2E-VarNet                      & \secondbest{35.40}                         & \secondbest{0.957}   & \secondbest{33.48}                         & \secondbest{0.945}                                   & \secondbest{32.79}                         & \secondbest{0.936}                   & \secondbest{31.02}                         & \secondbest{0.913} &\cellcolor{almond}\best{32.59}                         & \cellcolor{almond} \best{0.919}                           \\

\midrule
ShaRP           & \best{37.59}                         &\best{0.963}  & \best{34.66}                         &\best{0.949} &    \best{33.42}                         &\best{0.940} & \best{31.53}                         &\best{0.924}   &\cellcolor{almond} \secondbest{32.37}                         &\cellcolor{almond}\secondbest{0.907}                  \\
\bottomrule   
\end{tabular}


\caption{Quantitative comparison of ShaRP with task-specific baselines trained on the $8 \times$ uniform mask. Baselines perform well in matched settings (highlighted in the table) but show a significant drop under mismatched conditions. In contrast, ShaRP remains robust, handling both matched and mismatched scenarios effectively.}
\label{tab:mismatch_mri}

\end{table}

\clearpage
\section{Additional visual results for SISR}
\label{Sup:Sec:discussion_sisr}

In this section, we present additional visual results to numerical comparisons for the SISR task.

\subsection{Additional visual results against baselines}
\begin{figure}[h]
  \centering
  \includegraphics[width=.965\textwidth]{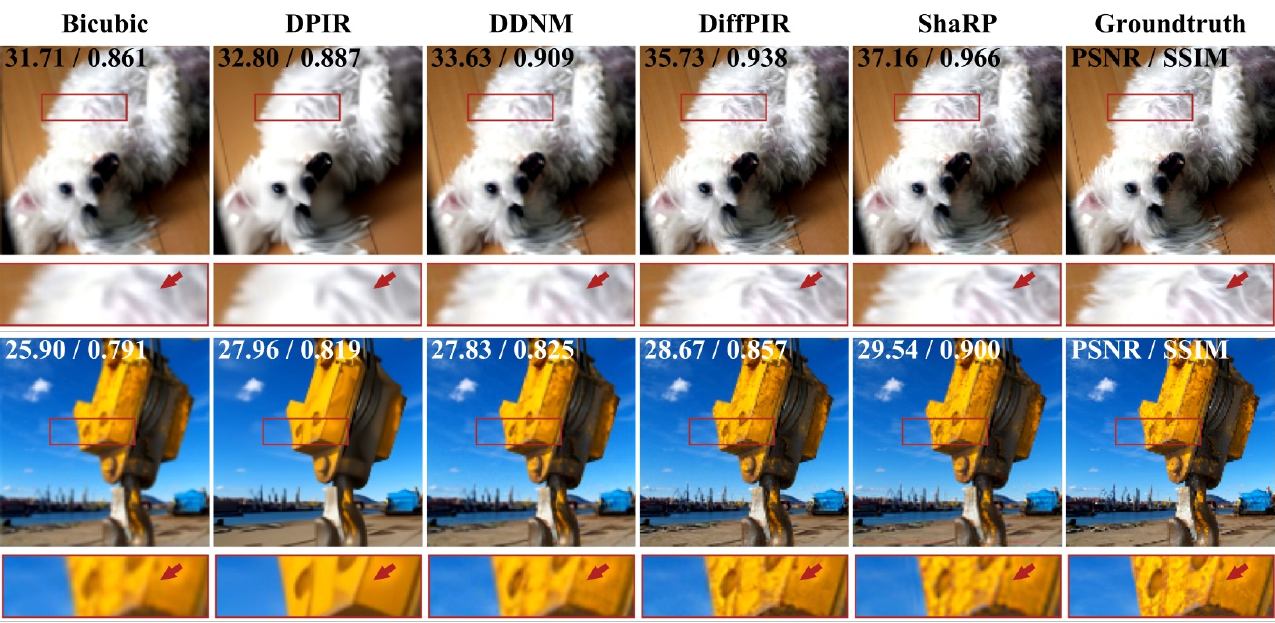}
  \caption{Visual comparison of ShaRP with several well-known methods on $2\times$ SISR with gaussian blur kernel with $\sigma = 1.5$. The quantities in the top-left corner of each image provide PSNR and SSIM values for each method. The squares at the bottom of each image visualize the zoomed area in the image.}
  \label{Sup:fig:sisr_1.25}
\end{figure}

\begin{figure}[h]
  \centering
  \includegraphics[width=.965\textwidth]{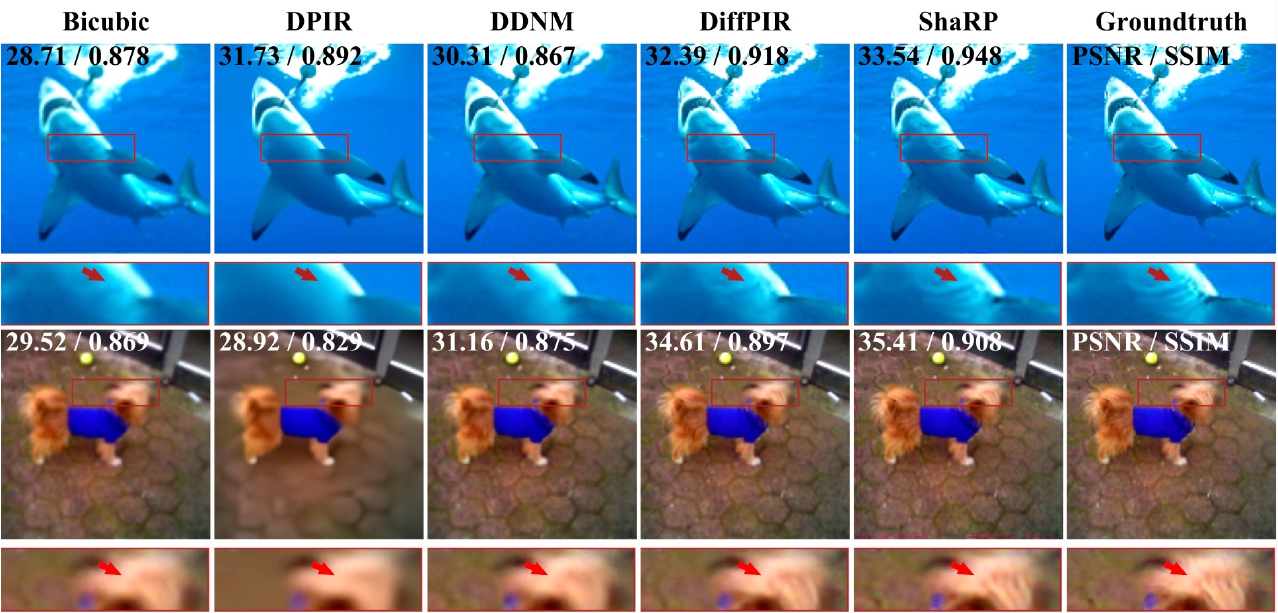}
  \caption{Visual comparison of ShaRP with several well-known methods on $2\times$ SISR with gaussian blur kernel with $\sigma = 1.5$. The quantities in the top-left corner of each image provide PSNR and SSIM values for each method. The squares at the bottom of each image visualize the zoomed area in the image.}
  \label{Sup:fig:sisr_1.5}
\end{figure}

As illustrated in Figure~\ref{Sup:fig:sisr_1.25} and Figure~\ref{Sup:fig:sisr_1.5}, ShaRP outperforms all baseline approaches under both blur kernel settings, achieving higher PSNR and SSIM values. Moreover, we maintain superior data consistency with the measurements while achieving enhanced perceptual quality. The use of an ensemble of deblurring priors enables our method to recover fine details at varying corruption levels, contributing to the improved performance.

\clearpage
\subsection{Additional visual results against DRP}

\begin{figure}[h]
  \centering
  \includegraphics[width=.765\textwidth]{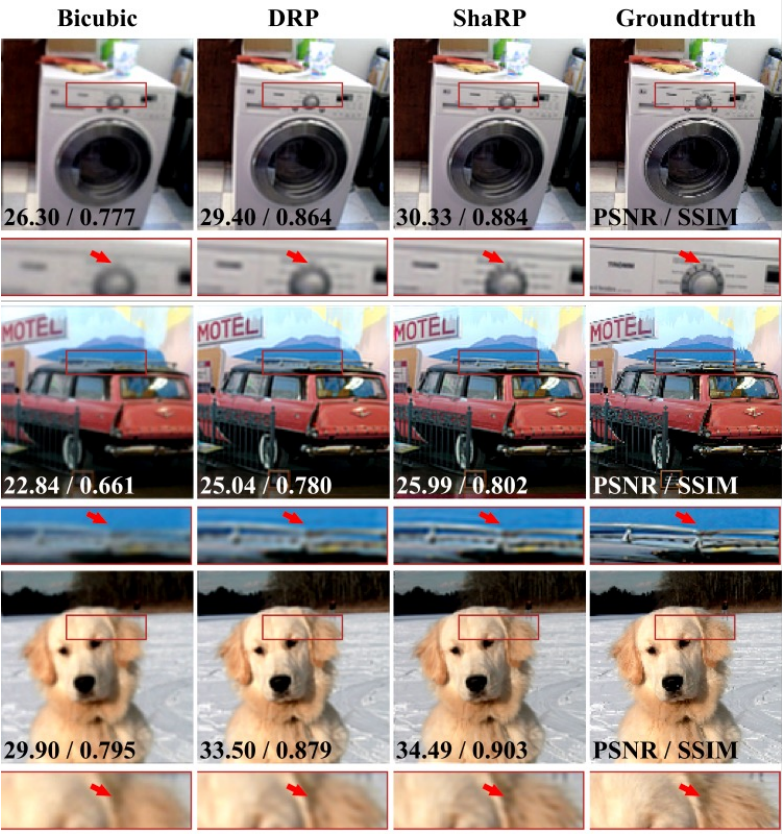}
  \caption{Visual comparison of ShaRP with DRP on $2\times$ SISR with gaussian blur kernel with $\sigma = 1.5$. The quantities in the bottom-left corner of each image provide PSNR and SSIM values for each method. The squares at the bottom of each image visualize the zoomed area in the image.}
  \label{Sup:fig:vs_drp_1.5}
\end{figure}

To further emphasize the necessity and advantages of using an ensemble of deblurring priors, as opposed to a fixed prior like in DRP~\citep{hurestoration}, we provide additional visual comparison results. As shown in Figure~\ref{Sup:fig:vs_drp_1.5}, ShaRP consistently recovers finer details, resulting in improved PSNR and SSIM scores, along with enhanced perceptual performance.

\end{document}